# Physical origin of current-induced switching angle shift in magnetic heterostructures

Xiaomiao Yin[1,3,4†], Guanglei Han[1,2†], Guowen Gong[1,2†], Jun Kang[3,4], Changmin Xiong[4,5], Lijun Zhu[1,2*]

1. *State Key Laboratory of Semiconductor Physics and Chip Technologies, Institute of Semiconductors, Chinese Academy of Sciences, Beijing 100083, China*
2. *Center of Materials Science and Optoelectronics Engineering, University of Chinese Academy of Sciences, Beijing 100049, China*
3. *Beijing Computational Science Research Center, Beijing, 100193, China*
4. *Department of Physics, Beijing Normal University, Beijing 100875, China*
5. *Key Laboratory of Multiscale Spin Physics, Ministry of Education, Beijing Normal University, Beijing 100875, China*

†These authors contributed equally. *ljzhu@semi.ac.cn

**Accurate quantification of the spin-orbit torques (SOTs) is critical for the identification and applications of new spin-orbitronic effects. One of the most popular techniques to qualify the SOTs is the "switching angle shift", where the applied direct current was assumed to shift, via domain wall depinning during the anti-domain expansion, the switching angle of a perpendicular magnetization in a linear proportion manner under a large rotating magnetic field. Here, we report that, for the most commonly employed perpendicular magnetization heterostructures in spintronics (e.g., those based on FeCoB, Co, and Co/Ni multilayers), the switching angle shift considerably misestimates the SOT within the domain wall depinning analysis of the slope of the linear-in-current scaling and may also have a non-zero residual value at zero direct current. Our experiments and simulations unveil that the switching angle shift is most likely dominated by the chiral asymmetric nucleation rather than the expansion of the anti-domains. The in-plane field from external magnet and current-induced SOTs lower the perpendicular nucleation field and thus the required switching angle, ultimately leading to underestimation of the SOTs by the domain wall depinning analysis. These results have advanced the understanding of magnetization switching of spintronic devices.**

Current-induced spin-orbit torques (SOTs) have attracted bloomed interest in the field of spintronics for the great potential in switching magnetization for fast, low-power memory and logic technologies [1-4]. Accurate quantification of the SOT efficiency ($\xi_{\rm DL}^j$) is critical not only for the usefulness evaluation of new magnetic heterostructures but also for the exploration of in-depth mechanisms of SOTs [5-12] and magnetization switching [14,15]. One of the most popular SOT analyses for magnetic heterostructures with perpendicular magnetic anisotropy (PMA) is the current-induced "switching angle shift" [15-27], which collects the Hall resistance ($R_{\rm H}$) as a function of the elevation angle ($\beta$, close to 0° or 180°) of a large magnetic field ($H_{xz}$) relative to the applied direct current ($I$, $x$ direction) that generates a damping-like SOT field ($H_{\rm DL}$) via the spin Hall effect (Fig. 1a-c). Assuming a switching mechanism of the conventional chiral domain wall depinning [28-32], the switching angle shift analysis predicts

$$H_{\rm DL} \approx H_{xz}\Delta\beta, \qquad (1)$$

where the switching angle shift is defined as $\Delta\beta \equiv (\beta_\downarrow+\beta_\uparrow)/2$ for $\beta \to 0°$ and $180° - (\beta_\downarrow+\beta_\uparrow)/2$ for $\beta \to 180°$, with $\beta_\downarrow$ and $\beta_\uparrow$ being the switching angles for the down-up and up-down switching. Thus, the switching angle shift model expects $\Delta\beta$ to scale inversely with $H_{xz}$ for a given current density ($j$) in the spin current-generating layer, linearly with $j$ for a given $H_{xz}$, and to be exactly zero at zero current density ($j = 0$). With the $\Delta\beta/j$ ratio, $\xi_{\rm DL}^j$ can be estimated as [28-32]

$$\xi_{\rm DL}^j = (4e/\pi\hbar)\,\mu_0 M_{\rm s}\,t_{\rm FM}\,H_{xz}\Delta\beta/j \qquad (2)$$

for the domain wall depinning switching. Here, $e$ is the elementary charge, $\hbar$ the reduced Planck's constant, $\mu_0$ the permeability of vacuum, $t_{\rm FM}$ the layer thickness of the magnetization, $M_{\rm s}$ the saturation magnetization.

Since 2012, current-induced switching angle shift has been widely employed and closely related to the evolution of the spintronics frontiers by evaluating the efficiencies and anisotropies of the SOTs of transverse spins, out-of-plane SOT fields, the Dresselhaus and Rashba effective fields in a variety of magnetic heterostructures, ranging from heavy metal/ferromagnet bilayers [21], single-layer ferromagnetic semiconductors (e.g., GaMnAs)[23], synthetic antiferromagnets (SAFs)[15-19], collinear antiferromagnets (e.g., Fe$_2$O$_3$ [22]), non-collinear antiferromagnets (e.g., Mn$_3$Sn [20]), van der Waals ferromagnets (Fe$_3$GeTe$_2$ [24]), spin valve trilayers [25], etc. Considerable claims have also been made from such switching angle shift analyses, including anisotropic self-induced SOTs in single-layer (Ga,Mn)(As,P)[27] and dramatic enhancement of the SOT efficiencies of transverse spins by magnetization compensation in SAFs [15](which is inconsistent with the fundamental physics that the competing spin relaxation mechanism of spin-orbit scattering that scatters spins into the lattice should the diminish the SOTs at full compensation [7]). However, it was barely verified as to whether or not domain wall depinning, the base of the switching angle shift analysis, was indeed the dominant switching mechanism for the practical magnetic heterostructures.

Recently, experiments [27,33-38] have revealed that domain wall depinning, both quantitatively and qualitatively, fails to describe the switching current density of perpendicular magnetization in practical samples. The domain wall depinning analysis of the critical switching current density is found to yield tens of times misestimation of the SOT efficiency [33,34]. The domain wall depinning cannot explain the strongly asymmetric magnetization switching driven by a magnetic field or by an in-plane current [35-38]. These advances suggest an urgent need to reevaluate the accuracy and impact of the switching angle shift technique and unveil the in-depth switching mechanisms.

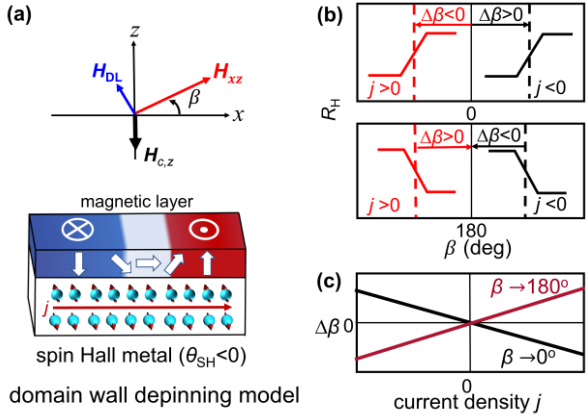

Fig. 1. Principle of the switching angle shift analysis. (a) Effective magnetic fields during spin-orbit torque switching via domain wall depinning in spin Hall metal/ferromagnet bilayer (for the case of a negative spin Hall ratio $\theta_{SH}$). (b) Hall resistance vs the elevation angle ($\beta$) of the applied magnetic field near 0° and 180°, indicating the sign of the switching angle shift ($\Delta\beta$). (c) Expected dependence of $\Delta\beta$ on direct current density for $\beta\to 0°$ and for $\beta\to 180°$.

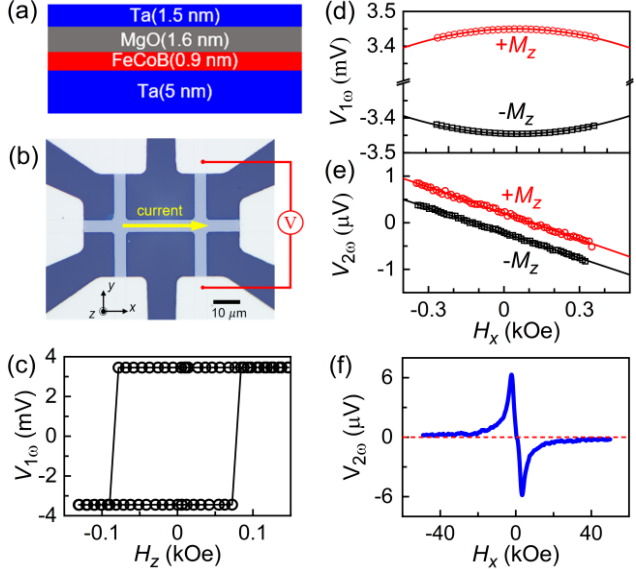

Fig. 2. Harmonic Hall voltage analysis of the Ta/FeCoB device. (a) Schematic of the magnetic stack, (b) Optical microscope image of the Hall bar device and the measurement coordinate. (c) First harmonic Hall voltage vs perpendicular magnetic field, (d) First and (e),(f) second harmonic Hall voltages vs the longitudinal magnetic field. $E$ = 62.7 kV/m. The data in (e) are shifted for clarity.

Here, we demonstrate that the switching angle shift analysis breaks down both qualitatively and quantitatively for magnetic heterostructures commonly used in spintronics. Experiments indicate that $\Delta\beta$ of magnetic bilayers typically includes a significant current-irrelevant contribution and a linear-in-current contribution that yields several times misestimation of $\xi_{DL}^j$ if the switching angle shift analysis was applied.

For this work, we first sputter-deposited a prototype magnetic stack of Ta(5)/FeCoB(0.9)/MgO(1.6)/Ta(1.5) (Fig. 2a), numbers are layer thicknesses in nm, FeCoB = $Fe_{60}Co_{20}B_{20}$) and several control samples (see below) on thermally oxidized silicon substrates at room temperature.

The saturation magnetization of the FeCoB layer is ≈1260 emu/cm$^3$ as measured using a superconducting quantum interference device. The samples are patterned into 5 μm×60 μm Hall bar devices (Fig. 2b) by photolithography and argon ion milling for transport measurements. The Ta/FeCoB devices are annealed at 210 °C for 20 min to enhance PMA. The resistivity ($\rho_{xx}$) of the Ta layer is 220 μΩ cm as calibrated from the conductance enhancement of the Ta(5)/FeCoB(0.9)/MgO(1.6)/Ta(1.5) relative to the control stack Ta(1)/FeCoB (0.9)/MgO(1.6)/Ta(1.5).

We characterize the magnetic anisotropy and the SOTs of the samples using the well-recognized harmonic Hall voltage measurement in the low-field macrospin regime [5,8,39]. The in-phase first and out-of-phase second harmonic Hall voltages, $V_{1\omega}$ and $V_{2\omega}$, are collected using a lock-in amplifier as a function of the magnetic fields while a sinusoidal electric field of $E$ = 62.7 kV/m is applied onto the 60 μm-long Hall bar. As revealed by the first harmonic Hall voltage hysteresis driven by the perpendicular magnetic field ($H_z$) in Fig. 2c, the Ta/FeCoB device exhibits good PMA, sharp magnetization switching, large perpendicular coercivity ($H_c$) of ≈80 Oe, and the high anomalous Hall voltage ($V_{AH}$) of 3.45 mV. As shown in Fig. 2d-e, the $V_{1\omega}$ and $V_{2\omega}$ data of the device are parabolic and linear functions of the longitudinal magnetic field ($H_x$), respectively, indicating that the device behaviors as a rigid macrospin when tilted slightly from its easy axis by the in-plane field $H_x$ of $\leqslant$0.3 kOe (the same is true for the data measured as a function of the transverse field $H_y$). This allows accurate determination of the PMA field ($H_k$) and $H_{DL}$ from the low-field harmonic Hall voltage analysis based on the macrospin model [8], i.e.,

$$V_{1\omega} = V_{AH}(1-H_x^2/2H_k^2), \quad (3)$$

$$H_{DL} = -2\frac{\partial V_{2\omega}}{\partial H_x}\bigg/\frac{\partial^2 V_{1\omega}}{\partial H_x^2} - 2H_k V_{ANE,z}/V_{AHE}. \quad (4)$$

where $V_{ANE,z}$ is the anomalous Nernst voltage that we measure from the $V_{2\omega}$ value at very high positive $H_x$ in Fig. 2f. $H_k$ is 2880 Oe as estimated from the fit of the data to Eq. (3) in Fig. 2d. With $H_{DL}$ = 8.8±0.2 Oe estimated using Eq. (4) and $j = E/\rho_{xx}$, the efficiency of the dampinglike SOT is then estimated as $\xi_{DL}^j$ = -0.105±0.001 for the Ta/FeCoB device following the macrospin relation

$$\xi_{DL}^j = (2e/\hbar)\mu_0 M_s t_{FM} H_{DL}/j. \quad (5)$$

Next we perform the switching angle shift measurement on the same Ta/FeCoB device. As shown in Fig. 3a, we sweep the elevation angle ($\beta$) of the magnetic field ($H_{xz}$) with a fixed magnitude of 2.5 kOe (greater than $H_k$) near the positive (+x direction, $\beta\to 0°$) and negative (-x direction, $\beta\to 180°$) current directions and determined the shift $\Delta\beta$ of the switching angle loop, under different bias direct currents ($I$). In Fig. 3b we summarize $\Delta\beta$ of the Ta/FeCoB device as a function of the total direct current for $H_{xz}$=2.5 kOe. As expected by the switching angle shift model (Fig. 1c), $\Delta\beta$ does show a good linear scaling with $I$, with a slope of $\partial(\Delta\beta)/\partial I$ =-0.19°/mA. However, with the slope of the linear scaling, the domain wall depinning model (Eq. (2)) only yields a $\xi_{DL}^j$ of -0.037 for the Ta/FeCoB, which is about 3 times smaller than its true value (-0.105 as estimated from the harmonic Hall voltage analysis). This deviation is consistent with our previous reports that domain wall depinning analysis of the switching current of thin-film and van der Waals magnets may misestimate the torque efficiency by up to tens of times [33,34]. More strikingly, the linear fit of the $\Delta\beta$ vs $I$ data also yields a nonzero intercept (Fig. 3b), indicating a nonzero perpendicular effective magnetic field at zero direct current. This observation cannot be explained by the domain wall depinning model [3,28-32].

In Fig. 3c, we further plot the $\Delta\beta$ values measured for different $H_{xz}$ and the separated in-plane field-antisymmetric ($\beta_0$) and in-plane field-symmetric components ($\Delta\beta$-$\beta_0$). $\beta_0$ is irrelevant to the sign and magnitude of the direct current in the Hall-bar device and thus cannot be attributed to any SOTs of spins or any thermal effects. $\Delta\beta$-$\beta_0$ switches sign upon reversal of the current direction and scales inversely with $H_{xz}$ only at high fields but not at lower fields. Both the presence of the field-antisymmetric $\beta_0$ and the low-field deviation of field-symmetric $\Delta\beta$-$\beta_0$ from the inverse field scaling (Fig. 3c) disagree with the switching angle shift model.

We emphasize that the breakdown of the switching angle shift technique is universal rather than just specific to the Ta/FeCoB devices. We have experimentally verified that the switching angle shift technique misestimates the spin-orbit torques typically by more than 2 times for a variety of other magnetic devices with different spin Hall metals and magnetic layers (e.g., Pt/Co, $Pt_{75}Ti_{25}$/[Co/Ni]$_3$, and $Au_{25}Pt_{75}$/Co in Table I). The breakdown of the switching angle shift technique cannot be attributed to the presence of any perpendicularly polarized spin currents [40] or perpendicular Oersted field [41], the generation of which requires a nonzero direct current and electric asymmetries within the device. The coherent rotation model is also unlikely to explain the switching of these Hall-bar devices (e.g., the gradual switching in Fig. 4a).

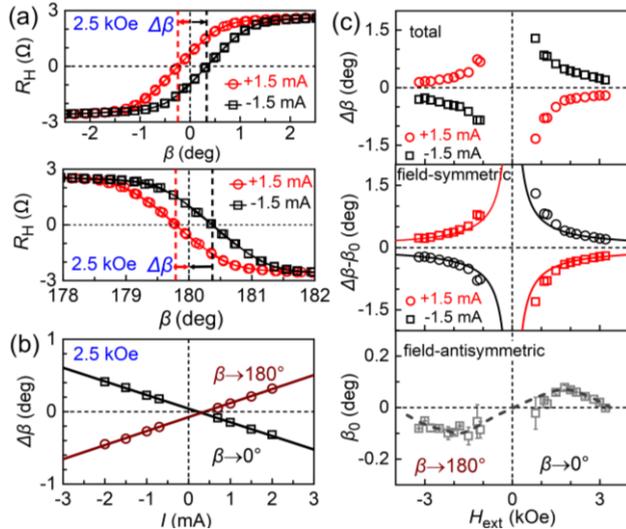

**Fig. 3. Switching angle shift analysis**. (a) Hall resistance vs the magnetic field angle measured at a total current of $I = \pm 1.5$ mA and a magnetic field of $H_{xz} = 2.5$ kOe. (b) Switching angle shift ($\Delta\beta$) vs current measured at $\beta$ around 0° or 180° and magnetic field of 2.5 kOe, the solid lines represent the best linear fits. Dependences on $H_{xz}$ of (c) $\Delta\beta$, $\Delta\beta$-$\beta_0$, and $\beta_0$. The solid black and red curves plot the best inverse field fits of the data, while the dashed dark gray curve is to guide the eyes.

We show below that the breakdown of the switching angle shift analysis arises because its base assumption, i.e., the domain depinning model, fails to capture the major characteristics of magnetization switching. The domain depinning assumption includes four characteristics [3,32,35,41-44], i.e., (i) the same current magnitudes for the SOT switching of perpendicular magnetization under a given in-plane magnetic field $H_x$, (ii) no switching of perpendicular magnetization by any in-plane magnetic fields, (iii) a $1/\cos\theta$ scaling of the switching magnetic field (i.e., $H_{sw} = H_{c,z}/\cos\theta$) at different polar angles ($\theta$), and (iv) a constant perpendicular coercivity ($H_{c,z}$) under different in-plane fields. Experimentally, the current switching hysteresis is shifted to the left side under a negative $H_x$ but to the right side under a positive $H_x$ (Fig. 4a). The Ta/FeCoB device exhibits sharp switching driven solely by $H_x$ in Fig. 4b. As shown in Fig. 4c, the switching field collected by sweeping the amplitude of the $H_{xz}$ field at fixed polar angles ($\theta$) also remarkably deviates near $\theta \approx \pm 90°$ (in-plane directions) from $1/\cos\theta$ scaling. In addition, the perpendicular coercivity $H_{c,z}$ also decreases with increasing $H_x$ towards zero (Fig. 4d). Thus, none of the characteristics of the domain wall depinning switching is fulfilled in the studied samples.

These switching characteristics of the practical samples are highly suggestive of a coercivity reduction and a chiral perpendicular effective magnetic field induced by the in-plane magnetic field, which we reproduce below using micromagnetic simulations with field-driven nucleation but without domain wall depinning (zero pinning field). Following the long-range intralayer Dzyaloshinskii–Moriya interaction (DMI) model that a chiral perpendicular magnetic field can arise from the collective chiral coupling of magnetic domains with anisotropy fluctuations [35], we simulate in Fig. 4e-g the switching process of a 500 nm×500 nm×1 nm PMA device (with a unit cell size of 1 nm×1 nm ×1 nm) that hosts a weaker-PMA 30 nm-in-diameter circular region near the center of the right edge (marked by a green dot in Fig. 4e) as a function of the in-plane magnetic field. Other parameters include the exchange stiffness $A = 0.8$ μerg/cm [45,46], $M_s = 1260$ emu/cm$^3$, the DMI constant $D = 0$ and 1.5 erg/cm$^2$, and $H_k = 2.75$ kOe for the weak PMA region and 2.80 kOe for the rest region (this is a rather weak anisotropy variation, less than 1.8%).

As shown in Fig. 4e, in the presence of both DMI and anisotropy fluctuations, the down-up switching (driven by a positive $H_z$) and up-down switching (driven by a negative $H_z$) undergo asymmetric nucleation (different nucleation fields) and propagation processes due to the $H_x$-trigged DMI coupling of the strong and weak PMA regions. Here, we stress that, while critical for the asymmetry of the switching, the tiny anisotropy fluctuations, as is usually the case for high-quality samples, is too weak to affect the location of anti-domain nucleation such that the anti-domains always nucleate at the center of left or right edges due to the interfacial DMI[47]. Macroscopic consequences are that the in-plane magnetic field induces a rapid decrease of $H_{c,z}$ towards zero in the presence/absence of DMI (Fig. 4g) and a small hysteresis shift (Fig. 4f) that varies anti-symmetrically and non-monotonically with $H_x$ only in the presence of a perpendicular intralayer DMI field ($H_{DMI}^z$, the red dashed curve in Fig. 4g). Such chiral asymmetric nucleation effect reasonably explains the seemingly puzzling experimental results in Fig. 3c and Fig. 4a-d. Switching of the PMA device by the in-plane field can be expected when $H_{c,z}$ becomes zero or less than the intralayer DMI field. These results suggest a critical role of chiral asymmetric nucleation in the switching of perpendicular magnetization. It is also important to note that, in the absence of the DMI effects (i.e., at zero DMI and/or zero anisotropy fluctuations), the nucleation field also decreases rapidly with the in-plane magnetic field but in a symmetric manner (Fig. 4g).

**Table 1**. Comparison of the spin-orbit torque efficiencies ($\xi_{DL}^j$) estimated from the harmonic Hall voltage and switching angle shift analyses.

| samples | $H_k$ (kOe) | $\xi_{DL}^j$ harmonic Hall voltage | switching angle shift ($H_{xz}$ = 2.5 kOe) | ratio |
|---|---|---|---|---|
| Ta 5/FeCoB 0.9 | 2.9 | -0.105±0.001 | -0.037 | 3 |
| Pt 4/Co 1 | 10.6±0.1 | 0.219± 0.003 | 0.029± 0.003 | 7 |
| Au$_{25}$Pt$_{75}$ 4/Co 0.64 | 6.6±0.1 | 0.305±0.013 | 0.153 | 2 |
| Pt$_{75}$Ti$_{25}$ 2/[Co 0.25/Ni 0.25]$_3$ | 4.1 | 0.18 | 0.064 | 2.8 |

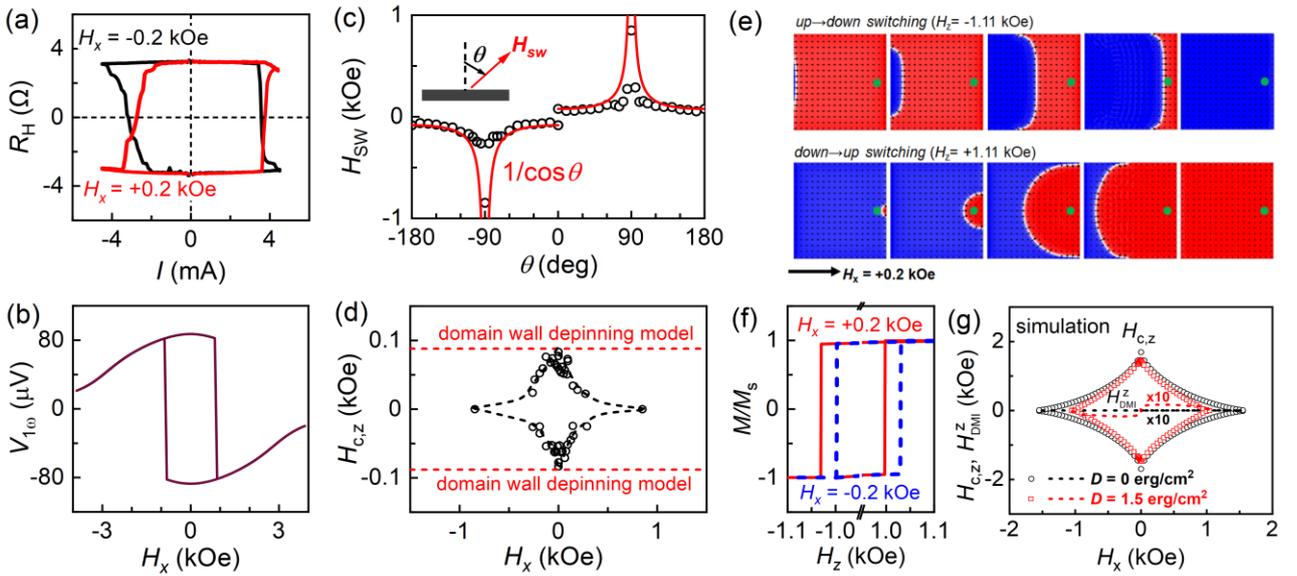

**Fig. 4. Deviation from the domain wall depinning switching model**. (a) Hall resistance vs the total in-plane current under a longitudinal magnetic field of $H_x$=±200 Oe, revealing a sizable shift induced by the in-plane magnetic field. (b) First harmonic Hall voltage vs $H_x$, revealing sharp switching of the perpendicular magnetization by the in-plane magnetic field. (c) Dependence of the total switching field on its polar angle ($\theta$), the red curve plots of the 1/cos$\theta$ fit, indicating strong deviation of the data from domain wall depinning at around ±90°. (d) Perpendicular coercivity ($H_{c,z}$) vs $H_x$. The red dashed lines indicate the predicted $H_{c,z}$ values from the domain-wall depinning switching model. Small electric field of $E$ = 1.57 kV/m for measurement of (b)-(d) to assure minimal influence of Joule heating. Micromagnetic simulation results of (e) the z field-driven up-down and down-up switching process ($H_x$ = +0.2 kOe, $D$ = 1.5 erg/cm$^2$, the green dot near the right edge marks the location of a weak PMA region), (f) hysteresis loops ($H_x$ = ±0.2 kOe, $D$ = 1.5 erg/cm$^2$), (g) Dependences on $H_x$ of $H_{c,z}$ (circles and squares) and the intralayer DMI field $H_{DMI}^z$ (dashed lines, multiplied by 10 for clarity) for a PMA device with $D$ = 0 (black) and 1.5 erg/cm$^2$ (red).

In conclusion, we have demonstrated from experiments and micromagnetic simulations that the popular "switching angle shift" technique can yield considerable misestimate for the spin-orbit torques of magnetic heterostructures. We find that the inaccuracy of the "switching angle shift" analysis arises because its base assumption, the simplifying domain wall depinning model, has not captured the remarkable, widespread in-plane magnetic field effects that arise most likely from chiral anisotropic nucleation. These findings suggest that switching angle shift analysis is not a universal technique for spin-orbit torque characterization but should be limited to the samples strictly fulfilling the conventional domain wall depinning model, the latter can be experimentally verified by its four characteristics that we specified above. We note that it is unrealistic to expect a universal analytical description of the relation of switching angle shift and the SOTs, e.g., via a simple modification of Eq. (1), for various magnetic samples (e.g., magnetic single-layers, heavy metal/magnet bilayers, or SAF trilayers) with the switching dominated by or involved with chiral asymmetric nucleation due to the considerable complications of the interplay of the magnetization switching, various DMI effects (i.e., bulk, intralayer, interlayer DMIs)[35,47-52], and anisotropy fluctuations[35]. Moreover, this work also provides the chiral asymmetric nucleation as a potential microscopic mechanism for understanding the various magnetization switching puzzles, such as in-plane field induced coercivity variation (Fig. 4d), non-linear current dependence of the switching field shift [53-55], large discrepancies of the spin-torque efficiencies from switching angle/field shift analysis and others [56], intralayer DMI fields [35], asymmetric switching (Fig. 4f), and misestimation of the SOT efficiency from domain wall depinning analysis of the switching current [33]. This work will stimulate further efforts towards more precise, in-depth understanding of the magnetization switching (the central topic of spintronics) and recheck of the considerable literature claims relying on switching angle shift and domain wall depinning analysis.

This work is supported partly by the Beijing Natural Science Foundation (Z230006), by the National Key Research and Development Program of China (2022YFA1204000), partly by the National Natural Science Foundation of China (12274405, 12304155, 12393831).

**Reference**
[1] L. Liu, C.-F. Pai, Y. Li, H. W. Tseng, D. C. Ralph, R. A. Buhrman, Spin-torque switching with the giant spin Hall effect of tantalum, Science 336, 555 (2012).
[2] I. M. Miron, K. Garello, G. Gaudin, P.-J. Zermatten, M. V. Costache, S. Auffret, S. Bandiera, B. Rodmacq, A. Schuhl, P. Gambardella, Perpendicular switching of a single ferromagnetic layer induced by in-plane current injection. Nature (London) 476, 189 (2011).
[3] L. Zhu, Switching of Perpendicular Magnetization by Spin-Orbit Torque. Adv. Mater. 35, 2300853 (2023).
[4] Z. Luo, A. Hrabec, T. P. Dao, G. Sala, S. Finizio, J. Feng, S. Mayr, J. Raabe, P. Gambardella, L. J. Heyderman, Current-driven magnetic domain-wall logic, Nature, 579, 214-218 (2020).
[5] L. Zhu, D. C. Ralph, R. A. Buhrman, Highly efficient spin current generation by the spin Hall effect in $Au_{1-x}Pt_x$, Phys. Rev. Appl. 10, 031001 (2018).
[6] L. Zhu, L. Zhu, S. Shi, M. Sui, D.C. Ralph, R.A. Buhrman, Enhancing Spin-Orbit Torque by Strong Interfacial Scattering From Ultrathin Insertion Layers, Phys. Rev. Appl. 11, 061004 (2019).
[7] L. Zhu, D.C. Ralph, Strong variation of spin-orbit torques with relative spin relaxation rates in ferrimagnets, Nat. Commun. 14, 1778 (2023).
[8] L. Zhu, J. Li, L. Zhu, X. Xie, Boosting Spin-Orbit-Torque Efficiency in Spin-Current-Generator/Magnet/Oxide Superlattices, Phys. Rev. Appl. 18, 064052 (2022).
[9] J. Han, A. Richardella, S.A. Siddiqui, J. Finley, N. Samarth, L. Liu, Room-Temperature Spin-Orbit Torque Switching Induced by a Topological Insulator, Phys. Rev. Lett. 119, 077702 (2017).
[10] M. DC, R. Grassi, J.-Y. Chen, M. Jamali, D. R.Hickey, D. Zhang, Z. Zhao, H. Li, P. Quarterman, Y. Lv, M. Li, A. Manchon, K. A. Mkhoyan, T. Low, J.-P. Wang, Room-temperature high spin–orbit torque due to quantum confinement in sputtered $Bi_xSe_{(1-x)}$ films, Nat. Mater. 17, 800–807 (2018).
[11] L. Zhu, L. Zhu, R. A. Buhrman, Fully spin-transparent magnetic interfaces enabled by insertion of a thin paramagnetic NiO layer, Phys. Rev. Lett. 126, 107204 (2021).
[12] H. Wu, P. Zhang, P. Deng, Q. Lan, Q. Pan, S. A. Razavi, X. Che, L. Huang, B. Dai, K. Wong, X. Han, and K. L. Wang, Room-Temperature Spin-Orbit Torque from Topological Surface States, Phys. Rev. Lett. 123, 207205 (2019).
[13] Z. Yan, Z. Li, L. Zhu, X. Lin, L. Zhu, Linear Enhancement of Spin-Orbit Torques and Absence of Bulk Rashba-Type Spin Splitting in Perpendicularly Magnetized $[Pt/Co/W]_n$ Superlattices, arXiv: 2412.18481 (2024).
[14] C. Zhang, S. Fukami, S. DuttaGupta, H. Sato, and H. Ohno, Time and spatial evolution of spin–orbit torque-induced magnetization switching in W/CoFeB/MgO structures with various sizes, Jpn. J. Appl. Phys. 57 04FN02 (2018).
[15] P. X. Zhang, L. Y. Liao, G. Y. Shi, R. Q. Zhang, H. Q. Wu, Y. Y. Wang, F. Pan, and C. Song, Spin-orbit torque in a completely compensated synthetic antiferromagnet, Phys. Rev. B 97, 214403 (2018).
[16] Y. Saito, S. Ikeda, H. Inoue, and T. Endoh, Charge-to-Spin Conversion Efficiency in Synthetic Antiferromagnetic System Using Pt–Cu/Ir/Pt–Cu Spacer Layers, IEEE Trans. Magn. 59, 1 (2023).
[17] H. Fan, M. Jin, B. Wu, M, Wei, J, Wang, Z. Shao, C. Yu, J. Wen, H. Li, W. Li, T. Zhou, Field-free switching and high spin–orbit torque efficiency in Co/Ir/CoFeB synthetic antiferromagnets deposited on miscut $Al_2O_3$ substrates, Appl. Phys. Lett. 122, 262404 (2023).
[18] H. Fan, M. Jin, Y. Luo, H. Yang, B. Wu, Z. Feng, Y. Zhuang, Z. Shao, C. Yu, H. Li, J. Wen, N. Wang, B. Liu, W. Li, T. Zhou, Field-Free Spin-Orbit Torque Switching in Synthetic Ferro and Antiferromagents with Exchange Field Gradient, Adv. Funct. Mater. 33, 2211953 (2023).
[19] Y. Saito, S. Ikeda, and T. Endoh, Enhancement of current to spin-current conversion and spin torque efficiencies in a synthetic antiferromagnetic layer based on a Pt/Ir/Pt spacer layer, Phys. Rev. B 105, 054421 (2022).
[20] J.-Y. Yoon, P. Zhang, C.-T. Chou, Y. Takeuchi, T. Uchimura, J.T. Hou, J. Han, S. Kanai, H. Ohno, S. Fukami, L. Liu, Handedness anomaly in a non-collinear antiferromagnet under spin–orbit torque, Nat. Mater. 22, 1106 (2023).
[21] L. Liu, O. J. Lee, T. J. Gudmundsen, D. C. Ralph, and R. A. Buhrman, Current-Induced Switching of Perpendicularly Magnetized Magnetic Layers Using Spin Torque from the Spin Hall Effect, Phys. Rev. Lett. 109, 096602 (2012).
[22] P. Zhang, C.-T. Chou, H. Yun, B. C. McGoldrick, J. T. Hou, K. A. Mkhoyan, and L. Liu, Control of Néel Vector with Spin-Orbit Torques in an Antiferromagnetic Insulator with Tilted Easy Plane, Phys. Rev. Lett. 129, 017203 (2022).
[23] M. Jiang, H. Asahara, S. Sato, T. Kanaki, H. Yamasaki, S. Ohya, and M. Tanaka, Efficient full spin–orbit torque switching in a single layer of a perpendicularly magnetized single-crystalline ferromagnet, Nat. Commun. 10, 2590 (2019).
[24] Y. Zhang, H. Xu, C. Yi, X. Wang, Y. Huang, J. Tang, J. Jiang, C. He, M. Zhao, T. Ma, J. Dong, C. Guo, J. Feng, C. Wan, H. Wei, H. Du, Y. Shi, G. Yu, G. Zhang,

X. Han, Exchange bias and spin–orbit torque in the Fe$_3$GeTe$_2$-based heterostructures prepared by vacuum exfoliation approach, Appl. Phys. Lett. 118, 262406 (2021).

[25] R. Q. Zhang, J. Su, J. W. Cai, G. Y. Shi, F. Li, L. Y. Liao, F. Pan, and C. Song, Spin valve effect induced by spin-orbit torque switching, Appl. Phys. Lett. 114, 092404 (2019).

[26] S. Park, K. J. Lee, K. Han, S. Lee, X. Liu, M. Dobrowolska, and J. K. Furdyna, Spin orbit torque switching of magnetization in the presence of two different orthogonal spin–orbit magnetic fields, Appl. Phys. Lett. 121, 112403 (2022).

[27] S. Park, K. J. Lee, S. Lee, X. Liu, M. Dobrowolska, and J. K. Furdyna, Spin–orbit torque switching in a single (Ga,Mn)(As,P) layer with perpendicular magnetic anisotropy, APL Mater. 9, 101102 (2021).

[28] A. Thiaville, S. Rohart, É. Jué, V. Cros, and A. Fert, Dynamics of Dzyaloshinskii domain walls in ultrathin magnetic films, EPL 100, 57002 (2012).

[29] D. Bhowmik, M. E. Nowakowski, L. You, O. Lee, D. Keating, M. Wong, J. Bokor, S. Salahuddin, Deterministic Domain Wall Motion Orthogonal To Current Flow Due To Spin Orbit Torque, Sci. Rep. 5, 11823 (2015).

[30] M. Baumgartner, K. Garello, J. Mendil, C. O. Avci, E. Grimaldi, C. Murer, J. Feng, M. Gabureac, C.. Stamm, Y. Acremann, S. Finizio, S. Wintz, J. Raabe, P. Gambardella, Spatially and time-resolved magnetization dynamics driven by spin–orbit torques, Nat. Nanotech. 12, 980 (2017).

[31] C.-F. Pai, M. Mann, A.J. Tan, G.S. Beach, Determination of spin torque efficiencies in heterostructures with perpendicular magnetic anisotropy, Phys. Rev. B, 93, 144409 (2016).

[32] O. J. Lee, L. Q. Liu, C. F. Pai, Y. Li, H. W. Tseng, P. G. Gowtham, J. P. Park, D. C. Ralph, R. A. Buhrman, Central role of domain wall depinning for perpendicular magnetization switching driven by spin torque from the spin Hall effect, Phys. Rev. B, 89, 024418 (2014).

[33] L. Zhu, D.C. Ralph, R.A. Buhrman, Lack of Simple Correlation between Switching Current Density and Spin-Orbit-Torque Efficiency of Perpendicularly Magnetized Spin-Current-Generator-Ferromagnet Heterostructures, Phys. Rev. Appl. 15, 024059 (2021).

[34] X. Lin, L. Zhu, Magnetization switching in van der Waals systems by spin-orbit torque, Mater. Today Electron. 4, 100037 (2023).

[35] Q. Liu, L. Liu, G. Xing, L. Zhu, Strongly asymmetric magnetization switching and programmable complete Boolean logic enabled by long-range intralayer Dzyaloshinskii-Moriya interaction, Nat. Commun. 15, 2978 (2024).

[36] G. Yu, P. Upadhyaya, K. L. Wong, W. Jiang, J. G. Alzate, J. Tang, P. K. Amiri, K. L. Wang, Magnetization switching through spin-Hall-effect-induced chiral domain wall propagation, Phys. Rev. B 89, 104421 (2014).

[37] W. Fan, J. Zhao, M. Tang, H. Chen, H. Yang, W. Lü, Z. Shi, X. Qiu, Asymmetric Spin-Orbit-Torque-Induced Magnetization Switching With a Noncollinear In-Plane Assisting Magnetic Field, Phys. Rev. Appl. 11, 034018 (2019).

[38] Q. Hao, G. Xiao, Giant spin Hall effect and magnetotransport in a Ta/CoFeB/MgO layered structure: A temperature dependence study, Phys. Rev. B 91, 224413 (2015).

[39] M. Hayashi, J. Kim, M. Yamanouchi, and H. Ohno, Quantitative characterization of the spin-orbit torque using harmonic Hall voltage measurements, Phys. Rev. B 89, 144425 (2014).

[40] Q. Liu, X. Lin, Z. Nie, G. Yu, L. Zhu, Efficient generation of out-of-plane polarized spin current in polycrystalline heavy metal devices with broken electric symmetries, Adv. Mater. 36, 2406552 (2024).

[41] Q. Liu, L. Zhu, Current-induced perpendicular effective magnetic field in magnetic heterostructures, Appl. Phys. Rev. 9, 041401 (2022).

[42] E. Kondorsky, On hysteresis in ferromagnetics, J. Phys. USSR **2**, 161 (1940).

[43] F. Schumacher, On the modification of the Kondorsky function, J. Appl. Phys. 70, 3184 (1991)

[44] L. Zhu, L. Zhu, Q. Liu, X. Lin, Giant coercivity, resistivity upturn, and anomalous Hall effect in ferrimagnetic FeTb, Phys. Rev. B 108, 014420 (2023).

[45] J. Kim, et al. Control of crystallization and magnetic properties of CoFeB by boron concentration, Sci. Rep. 12, 4549 (2022).

[46] M. Yamanouchi et al. Domain Structure in CoFeB Thin Films With Perpendicular Magnetic Anisotropy, IEEE Magnetics Letters 2, 3000304 (2011).

[47] S. Pizzini, J. Vogel, S. Rohart, L. D. Buda-Prejbeanu, E. Jué, O. Boulle, I. M. Miron, C. K. Safeer, S. Auffret, G. Gaudin, and A. Thiaville, Chirality-Induced Asymmetric Magnetic Nucleation in Pt/Co/AlO$_x$ Ultrathin Microstructures, Phys. Rev. Lett. 113, 047203 (2014).

[48] A. Fernandez-Pacheco et al. Symmetry-breaking interlayer Dzyaloshinskii-Moriya interactions in synthetic antiferromagnets, Nat. Mater. 18, 679 (2019).

[49] D. Han, et al. Long-range chiral exchange interaction in synthetic antiferromagnets, Nat. Mater. 18, 703 (2019).

[50] C. O. Avci, C. Lambert, G. Sala, P. Gambardella, Chiral coupling between magnetic layers with orthogonal magnetization, Phys. Rev. Lett. 127, 167202 (2021).

[51] L. Zhu, L. Zhu, X. Ma, X. Li, R. A. Buhrman, Critical role of orbital hybridization in Dzyaloshinskii-Mariya interaction of magnetic interfaces, Commun. Phys. 5, 151 (2022).

[52] L. Zhu, D. Lujan, X. Li, Discovery of strong bulk Dzyaloshinskii-Moriya interaction in composition-uniform centrosymmetric magnetic single layers, Sci. China-Phys. Mech. Astron. 67, 227511 (2024).

[53] T. Dohi, S. Fukami, and H. Ohno, Influence of domain wall anisotropy on the current-induced hysteresis loop shift for quantification of the Dzyaloshinskii-Moriya interaction, Phys. Rev. B 103, 214450 (2021).

[54] Y. Wu, T. Wu, H. Chen, Y. Cui, H. Xu, N. Jiang, Z. Cheng, Y. Wu, Field-free spin–orbit switching of canted magnetization in Pt/Co/Ru/RuO$_2$(101) multilayers, Appl. Phys. Lett. 126, 012401 (2025).

[55] Q. Yang, D. Han, S. Zhao, J. Kang, F. Wang, S.-C. Lee,

J. Lei, K.-J. Lee, B.-G. Park, H. Yang, Field-free spin–orbit torque switching in ferromagnetic trilayers at sub-ns timescales, Nat. Commun. 15, 1814 (2024)

[56] L. Zhu, D. C. Ralph, R. A. Buhrman, Maximizing Spin-Orbit Torque Generated by the Spin Hall Effect of Pt, Appl. Phys. Rev. 8, 031308 (2021).